\documentclass[conference]{IEEEtran}
\ifCLASSINFOpdf
  \usepackage[pdftex]{graphicx}
  \DeclareGraphicsExtensions{.pdf,.jpeg,.png}
\else
\fi

\begin{document}
  \title{The Challenges in SDN/ML Based Network Security : A Survey}

\author{\IEEEauthorblockN{Tam Nguyen}
\IEEEauthorblockA{
North Carolina State University\\
tam.nguyen@ncsu.edu\\
https://www.linkedin.com/in/tamcs/}
}

\maketitle
\begin{abstract}
Machine learning is gaining popularity in the network security domain as many more network-enabled devices get connected, as malicious activities become stealthier, and as new technologies like Software Deﬁned Networking (SDN) emerge. Sitting at the application layer and communicating with the control layer, machine learning based SDN security models exercise a huge influence on the routing/switching of the entire Software Defined Network. Compromising the models is therefore a very desirable goal. Previous surveys have been done on either adversarial machine learning without the context of secure networking environment or the general vulnerabilities of SDNs without much consideration of the defending ML models. Through examination of the latest ML based SDN security applications and a good look at ML/SDN specific vulnerabilities accompanied by common attack methods on ML, this paper serves as a unique survey, making a case for more secure development processes of ML-based SDN security applications.
\end{abstract}


\IEEEpeerreviewmaketitle

\section{Introduction}
There is a significant gap between the amounts of connected devices and the number of cyber security professionals. Per U.S. Bureau of Labor Statistics \cite{USBureauofLaborStatistics2016}, a projected growth in cyber security jobs from 2014 to 2024 is 18\% while Cisco \cite{Forecast2017} predicted a 100\% increase in network-enabled devices, growing from 4 billions in 2016 to 8 billions units in 2021. Consequently, global data traffics will increase by at least 5 times. Furthermore, the emergence of Software Defined Network makes machine learning (ML) based network security solutions even more appealing. Usually, researchers design new ML-based security solutions and then use benchmark results to prove the models' accuracy. Such methodology alone is not attractive enough in the eyes of cybersecurity leaders due to the lack of information on how those solutions will fit into the bigger picture at their organizations. Other than accuracy, the cybersecurity leaders would also like to know about the model's projected maintenance costs, the model's ability to withstand abuses, the quality of source codes, the ways datasets were collected, and so on. Previous surveys have been done on either adversarial machine learning or the general vulnerabilities of SDNs but not both. For the first time, this paper aims for a complete picture regarding ML-based security solutions for SDN, addressing:\\
1. The latest landscape on ML enabled SDN security solutions\\
2. The vulnerabilities of the common ML models being used\\
3. The general ways the ML models can be attacked\\
4. What can be done to better develop ML-based SDN security solutions.\\
The paper recognizes that ML models deployed in a network detection/prevention system are not perfect. Hence there is always a possibility for attackers to manipulate and/or bypass the models. Once hackers can bypass the current state of the model, they may also be able to predict and bypass future states. From the beginning of their project, solution designers should pay special attentions to the threat model, the secure development processes, and so on.

\begin{figure*}
  \centering
  \includegraphics[width=15cm]{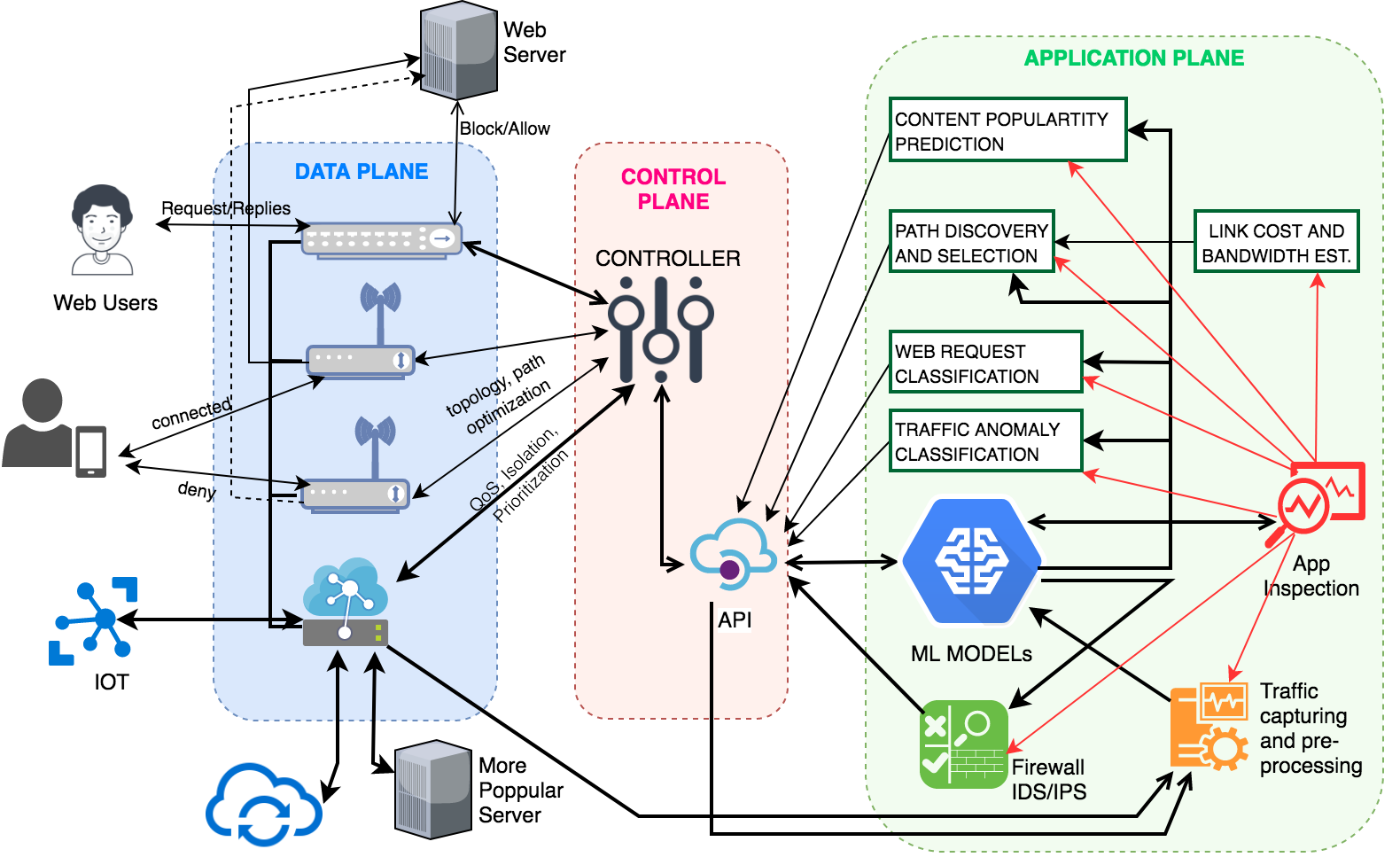}
  \caption{The Roles of ML Models in 2017 Software Defined Network IEEE Xplore Published Papers}
  \label{Figure:ML Roles in SDN}
\end{figure*}

\section{SDN based network security solutions}

\subsection{Background on SDNs}
Software Defined Networking was motivated by the inflexibility of traditional networks. Since administrators have to manually configure each main box (router/switch), changes do not happen fast and accurate enough. Additional factors like different manufacturers for network devices also make difficult to find the right professionals and to scale up the infrastructure when situations demand. The best example is the migration from IPv4 to IPv6. The process started several years ago and yet, we are still at it.\\
Fundamentally, SDNs differ from traditional networks on the key principle of separating network policies completely from network implementations \cite{Kreutz2013TowardsNetworks}. It consists of 4 pillars : 1. Separating control plane (CP) from data plane (DP)  2. Flow-based forwarding (instead of destination based) 3. NOS - network "operating system" - controls all the flow logics 4. Everything is programmable through APIs.  In fact, well-defined APIs between the planes are the indicator of a good SDN as they are separated into Southbound group (facing the data plane) and Northbound group (facing the application plane). OpenFlow is one example of such API. During normal operations, CP will dynamically modify the flow table which consists of 3 components: 1- A matching rule (based on fields like switch port, TCP ports, Vlan ID, MAC, Ethernet type) 2- Actions per matched rule/packets (forward, encapsulate, drop, send to processing pipelines) 3- Counters keeping track of matched packets to form per table/per flow/per port/per queue statistics. Some example of SDN controlers include NMS, NOX, POX, Floodlight, ProCel, Beacon, OpenDaylight, ProgrammableFlow, FlowVisor, OpenSketch, Procera, Ethane, Trema, SANE.\\
This centralized control design brings flexibility, simplicity, elasticity, promptness and many other benefits. It is important to note that while flow control is centralized, the CP can be physically distributed. Some SDN implementations are already at a very large scale production level such as the one being used by Google to connect its data centers across the globe. 

\subsection{SDN based network security solutions}
\subsubsection{Policy Enforcement}
Products like SANE \cite{Casado2006SANE:Networks} allow enforcement of simple and  natural access control policies at the link layer - independent from topologies while hiding topology and service information from those without the needs to know. Between a server B and client A, there are 3 steps in SANE's model. First, each party has to be authenticated with the controller. Second, B publishes to the controller what kind of service it is willing to provide and A requests from the controller what kind of service it needs. Finally, server double checks everything and establishes service accordingly. More than just access control, SDN also supports other diverse security policies relating to intrusion detection, virus scanning, protocol identification, etc. with linearly-increasing performance. One example is LiveSec \cite{Wang2012LiveSec:Networks} with its "interactive policy enforcement" in which network administrators can add/remove both rules and network security services easily with visual feedbacks. It is supported by a global policy table enforced by dynamic modification of MAC addresses. The benefits of SDN becomes more obvious when it comes to virtual and cloud environments. We can apply different sets of policies onto different types of VMs being created dynamically on the cloud, providing protection services similar to firewalls as well as elastic IP service. \cite{Stabler2012ElasticOpenFlow} Because the controller has a detailed overview of flows and flow paths on its network, enforcing address validation policies is efficient. \cite{Yao2011SourceArchitecture} For example, if there are B and C between the only path from A to D, a legitimate flow from A to D will have to leave its traces in A, B, C and D. With such knowledge, the controller can easily spot a spoofed flow A to D after figuring out that the flow did not travel through B and C.

\subsubsection{Denial of Service Mitigation}
The key point in protecting networks from being DOS'ed is the recognition of what flows of traffics are malicious as soon as possible. Because malicious packets are very similar to legitimate packets, the use of machine learning (ML) algorithms for automatic flow classification is not new. Traditional methods involve the pre-processing of traffic log files and captured packets in order to produce some basic statistics that the ML models can use. Such high overhead practice is not needed in SDN because counters are embedded within each network device in the data plane. The statics relating to forwarded traffics are always ready and can be queried by the control plane at anytime. A DOS detection loop contains 3 components: the Flow Collector, the Feature Extractor, and the Classifier. \cite{Braga2010LightweightNOX/OpenFlow} Once identified, malicious traffics should be dropped or forwarded to a null interface as being used in Remote Triggered Blackhole Routing Component (RTBH). The collateral damage is high with legacy RTBH due to the inflexibility of the trigger routers. SDN can allow a much more flexible RTBH, dropping the malicious flows while still maintaining benign flows. \cite{Giotis2014LeveragingNetworks} 

\subsubsection{Cloud Security}
In current PaaS offerings, everything follows the service model of which cycle involves (1) provision (2) bind, (3) unbind and (4) provision. Implementing cloud network security functions is therefore difficult. For one reason, the "wiring" between services can get really complicated and time consuming. For another reason, it is very difficult to reach to the network packet level from the cloud application level - all packet fields are hidden. Going to the rescue, SDN allows cloud network security services like IDS/IPS to: +manipulate traffics with minimal data copying +operate at both packet and request levels +conveniently generate callbacks via existing APIs +be chained with other network services easily.
\cite{Jamjoom2014DontMiddlepipes} Many high level routing algorithms can be designed and implemented in a way that guarantees all packets will be inspected by at least one security device.\cite{Shin2012CloudWatcher:Clouds}

\subsubsection{Topology Protection}
Protecting network topology is crucial in SDNs since all popular SDN controller are all subjected to network topology attacks \cite{Hong2015PoisoningCountermeasures}. Any changes to flow behaviors should be flagged for immediate remedy. Solutions can be flow-graph based \cite{Dhawan2015SPHINX:Networks} in which flow-graph of flows will be incrementally built and verified in real time. Verification is either deterministic (following the edges of graph) or probabilistic. Other solutions include VeriFlow \cite{Khurshid2013Veriflow:Time}, AvantGuard \cite{Shin2013AVANT-GUARD}, and FloodGuard \cite{Wang2015FloodGuard:Networks}.

\subsubsection{Others}
The latest research landscape of ML based security solutions for SDN is shown in figure \ref{Figure:ML Roles in SDN}. Other security solutions for SDNs can be found in the "Software-Defined Networking: A Comprehensive Survey" by Krewtzet et. al. \cite{Kreutz2015Software-DefinedSurvey}.

\begin{table*}[!t]
  \centering
  \begin{tabular}{ p{5cm}  p{2.2cm}  p{1cm}  p{8.1cm} }
    \hline
    \\
    PUBLICATION & FUNCTIONALITY & SUCCESS & ML METHODOLOGY \\
    \\
    \hline
    \\
    AMPS: Application aware multipath flow routing using machine learning in SDN  \cite{Pasca2017AMPS:SDN} & Dynamic multipath flow routing & 98\% & Supervised learning on 40-features dataset with C4.5 Decision Tree algorithm \\
	\\
    Dynamic attack detection and mitigation in IoT using SDN \cite{Bhunia2017DynamicSDN} & IoT monitoring and protection & 98\% & Support Vector Machine model was used on processed data passed down from a learning module in order to classify traffics. Both linear and non-linear kernel (RBF) were used.\\
    \\
    Analytics-Enhanced Automated Code Verification
for Dependability of Software-Defined Networks \cite{Jagadeesan2017Analytics-EnhancedNetworks} & Detect and prevent malicious SDN app behaviors & 99\% & It is a combination of automated code verification (Java Path Finder) with ML analytic model (Ckmeans), being independent of underlying network topology\\
    \\
    An applied pattern-driven corpus to predictive analytics in mitigating SQL injection attack & Prevent SQL injection in the cloud & 98\% & Two-Class Support Vector Machine and Two-Class Logistic Regression with linear kernel implemented on Microsoft Azure Machine Learning \\
    \\
    Forecasting and anticipating SLO breaches in programmable networks \cite{Bendriss2017ForecastingNetworks} & Cognitive SLA enforcement & 90\% & Protecting Service Level Objectives (availability, response time, throughput) using Long Short Term Memory Recurrent Neural Network (able to identify previously seen patterns in new distorted samples) \\
    \\
    Access point selection algorithm for providing optimal AP in SDN-based wireless network \cite{Lee2017AccessNetwork} & Pick best QoS, less backhaul conges- tion  & depends & Traffic classification deployed at AP, using C5.0 Decision Tree algorithm \\
    \\
    Content Popularity Prediction and Caching for ICN: A Deep Learning Approach With SDN  \cite{Liu2017ContentSDN} & Cache management & 75\% mean accuracy & Improving cach operations by predicting the popularity of content using Deep-Learning-based Content Popularity Prediction (Stacked Auto-Encoders + Softmax)  \\
    \\
    AWESoME: Big Data for Automatic Web Service Management in SDN \cite{Trevisan2018AWESoME:SDN} & Web traffic engineering & 90\% & Annotation module at network edge classifies flows in real time based on bag-of-domains (using Apache Spark), flow-to-domain, domain-2-service, and service-to-rule. \\
    \\
    Athena: A Framework for Scalable Anomaly Detection in Software-Defined Networks \cite{Lee2017Athena:Networks} & Network security development framework & depends & 11 ML models were made available as library, allowing developers to quickly develop network security applications that can perform real-time detection and responses. \\
    \\
   FADM: DDoS Flooding Attack Detection and Mitigation System in Software-Defined Networking \cite{Hu2017FADM:Networking} & framework for DOS prevention & depends & Feature extraction is entropy-based and attack detection is powered by Support Vector Machine  \\
    \\
    \hline \\
  \end{tabular}
  \caption{10 ML-based security solutions for SDNs since 2017}
  \label{table:MLsolutions}
\end{table*}

\subsection{Anomaly Detection Using Machine Learning}
Compared with signature based detection, anomaly detection using ML is more scalable, and more flexible.\cite{Sultana2018SurveyApproaches} All machine learning approaches follow the same general steps of identifying/building learning data sets, feature extraction and classification.Selecting the right dataset is crucial because ML models can only identify anomalies based on what it has known (trained with).The more organic, diverse and properly prepared dataset we have, the more accurate our models will be. Pre-processing steps usually involves mapping symbolic values to numeric values, data scaling, etc. In the feature extraction step, we pick the optimal number of features that will be used by the model to successfully categorize the classes that we want. Common methods are dimensional reduction (mapping more dimension variables into fewer ones), clustering (identifying groups of items with similar characteristics), statistical sampling, measuring and pick samples based on entropy and so on.For classification, there are three approaches: 1. Supervised learning (the models learn from labeled data and predict unknown cases) 2. Unsupervised learning (the models learn the fundamentals of unlabeled data and predict unknown cases) 3. Semi-supervised learning.Further details behind ML algorithms and methodologies can be found in numerous ML general surveys \cite{Kwon2017ADetection}.\\
In the domain of network security, SDN brings some unique advantages to the deployments of ML based network security solutions. For example, the centralized control sitting on the software layer with API access making it very convenient to develop ML softwares. Devices in SDN data plane has counters built-in and can provide statistical reports to application layer softwares upon requests. Table \ref{table:MLsolutions} provides us with the most notable SDN/ML research works indexed by IEEE Xplore from 2017 to MAR2018 (the time of this paper) from which we can see a broad range of ML based solutions and the incredible flexibility that SDN architecture can provide. It also shows that there is an on going strong interest from the research community in ML based network security solutions for SDN.

\section{Issues with ML Models in SDN security solutions}

In addition to existing problems within the protocols such as OpenFlow vulnerabitlies \cite{Benton2013OpenFlowAssessment}, ML-based security solutions for SDN also face issues with (1) hard to find organic training data; (2) a semantic gap between innitial work and practical real world deployments; (3) enormous variability in input data; (4) measuring and minimizing the cost of errors; and (5) other difficulties in evaluation \cite{Sommer2010OutsideDetection}.\\
A large portion of training data for ML-based SDN security applications is synthetic and is not realistic enough. Commonly used data sets include but not limiting to the University of New Brunswick ISCX 2012 Intrusion Detection, Evaluation Data Set, the CIC DOS Dataset, the KDD dataset, the ADFA-LD12 dataset, the UNSW-NB15 dataset, the WSN-DS dataset, and so on \cite{Sultana2018SurveyApproaches}. Because those datasets were developed by research institutes and made available to the public, ML-based solutions trained on those datasets alone can be out-maneuvered by adversaries who studied the same data. While some corporations have the capability to collect own training datasets from their existing networks, the fear of business secret, confidential communications, and employees personal identifiable information being leaked from such datasets really discourages them\\
Unlike Artificial Intelligence, a ML model works by recognizing the deviations from what it was trained on and because of problems with training datasets, ML models usually give plenty of false positives when being deployed in real-world environment. It is also difficult to interpret the overall results for actionable intelligence - a.k.a the Semantic Gap. For example, if the model's accuracy is 98\% on detecting some variations of social security numbers in http traffics, it does not give the administrators much information on how social security numbers are actually being leaked out of the network. The model does not know what it does not know, and it will skip leaks that are far different from the samples it was trained with.\\
Even when there is absolutely nothing wrong going on, performance of ML models in the real world is usually degraded because of a significantly higher volume of data, a much wider range of fluctuations, real world constraints, and so on. In short, while being good at classification tasks, ML models are not be able to recognize the context and logics behind real world situations \cite{Sultana2018SurveyApproaches}. Consequently, it is challenging to really evaluate a model. While there is no formal agreed standard on how to evaluate ML models and what metrics should be used, justification cannot be relied on the accuracy rate alone. In the following section, we will briefly examine the inherent limitations of common ML algorithms.

1) Artificial Neural Networks (ANNs) are fit for non-linear problems but tend to suffer from local minima leading to long learning time, and as the number of features increases, the longer it will take to learn. ANNS can reach deeper into lower network layer data and are able to detect some low/slow type attacks. While they can identify 100\% of the normal behavior, the amount of false alarms may sometimes reach 76\% depending on what kind of attacks were being executed \cite{Buczak2016}.

2) Bayesian network is a probabilistic directed acyclic graph type with nodes as variables and the edges as their relationships. Based on the relationships, a node can "walk" to another. At the end of the walk, a final probabilistic score is formed. Relationship links that have high true positive score will be verified and formed into rules. Therefore, a Bayesian network is proactive even in misuse mode. In a test of using model to label IRC-botnet generated data, the precision rate is 93\% with a false positive rate of 1.39\% (detecting fewer cases than some other models but generating less false alerts). In other tests, the reported precision rates are 89\%, 99\%, 21\%, 7\%, and 66\% for DoS, probe/scan, remote-to-local, user-to-root, and "other" classes of attacks respectively \cite{Buczak2016}.

3) Some popular clustering models are k-means, k-nearest neighbor, density-based spatial clustering of applications with noise (DBSCAN), etc. Explicit descriptions of classes are not required. However too many features may confuse the model and any imbalance in the feature set will negatively affect its decisions (the "curse of dimensionality"). Clustering can detect up to 80\% of unknown attacks. Some clustering models can be really accurate (98\%) in analyzing captured PCAP packages but the performance goes down when dealing with streaming data as false alarm rate may go up to 28\%.

4) Decision tree is a flow-chart like structure built on concepts of information gain/entropy where each node choose the best fit attribute to split current set of examples into subsets. Normally, decision trees provide high accuracy with simple implementation. It is not usually the case with larger trees where the model tends to favor attributes with more levels.

5) While Genetic Algorithm (GA) and Genetic Programming (GP) are most used Evolutionary Computation (EC) methods; Particle swarm optimization, Ant Colony Optimization, Evolution Strategies are also parts of the group. The main concept is based on the idea of "the strongest will prevail" and basic operators are selection, crossover, and mutation. Experiments with various attack types show that the average false alarm rate is very low. However, the sensitivity in detecting new attacks varies greatly (from 66\% to 100\%) depending on attack types.

6) Naive Bayes model calculates the final conditional probability of "attack" (or "normal") with a naive assumption that the used features are independent from each other. That assumption is the biggest limitation of this model. However, if the features are indeed independent from each other, naive Bayes can be very powerful thanks to its simple algorithm. It allows the model to be highly scalable and be used as an online classifier. In tests, model's accuracy of identifying data as "normal" is 97\% for DoS type attacks while only 9\% for remote-to-local type attacks. 
 
7) Supported Vector Machine (SVM) is a binary classification model by design. With a kernel such as linear, polynomial, Gaussian Radial Basis Function, or hyperbolic tangen; the model will try to draw a hyperplane that divides the feature space into two classes. Sometimes, when overlapping is unavoidable, slack variables will be added and each overlapping data point will be be assigned a cost value. The model is quite accurate but also shows limitations at identifying certain types of attacks such as user-to-root attack.Test  results show great variations in accuracy (from 65\% to 99.9\%) and sometimes, false negative rate can get really high (over 30\%) \cite{Buczak2016}.

\begin{table}[!t]
\centering 
\begin{tabular}{|c|c|c|c|c|} 
\hline
Papers & FPR (\%) & FNR (\%) & TPR (\%) & TNR (\%) \\ 
 \hline
BotMiner \cite{Gu2008BotMiner:Detectionb} & 0.1875 & --- & 96.82 & 81.25\\
 \hline
BotSniffer \cite{Gu2008BotSnifferTraffic} & 0.1600  & --- & 100.00  & 84.00\\
 \hline
N-gram \cite{Lu2011ClusteringSelection}  & 8.1400  & 1.64  & 98.36  & 91.86\\
 \hline
Unclean \cite{Collins2007UsingAddresses} & 1.2100 & 1.22  & 98.78  & 87.90\\
 \hline
FluXOR \cite{Passerini2008FluXOR:Networks} & 0.0000 & --- & --- & 100.00\\
 \hline
Tight \cite{Strayer2006DetectingControl} & 15.0400  & 2.49  & 75.10  & 84.96\\
 \hline
\end{tabular}
\newline
\caption{Accuracy-based performance metrics comparison}
\label{table:BotDetectionAccuracy}
\end{table}

\section{Attacking ML Models}
The definition of "attack" on ML models should be flexible and be focusing on the models' purposes rather than the models' functionalities. The accuracy rate is not the only thing adversaries can target. For example, adversaries can cause the models to produce true positives that are very close to false positives. Consequently, it causes burn-outs on the security analysts who are going to manually inspect those flags. "Attack" can also mean significantly increasing the time it takes for a ML model to make a specific decision. No matter what attack end-goals are, attackers must first have some really good insights about the targeted model. Thus, there is a strong motivation to clone/extract a security ML model. When performing model extraction, inputs are given to a trained model, end results (outputs) got harvested, and clone model learns from those input-output data pairs. While it appears that training the original model and cloning an existing model are quite similar, model cloning does not have to deal with broken or faulty data entries that delay or even mislead the learning process. Model cloning also does not have to deal with optimization issues such as local minima/maxima traps. In 2016, Tramer et. al. \cite{Tramer2016} proposed several methods to perform model extraction of several ML types but those methods were not weaponized. For example, the number of probes used was too high to be considered practical in targeting a protected environment.

\begin{figure*}
  \centering
  \includegraphics[width=15cm]{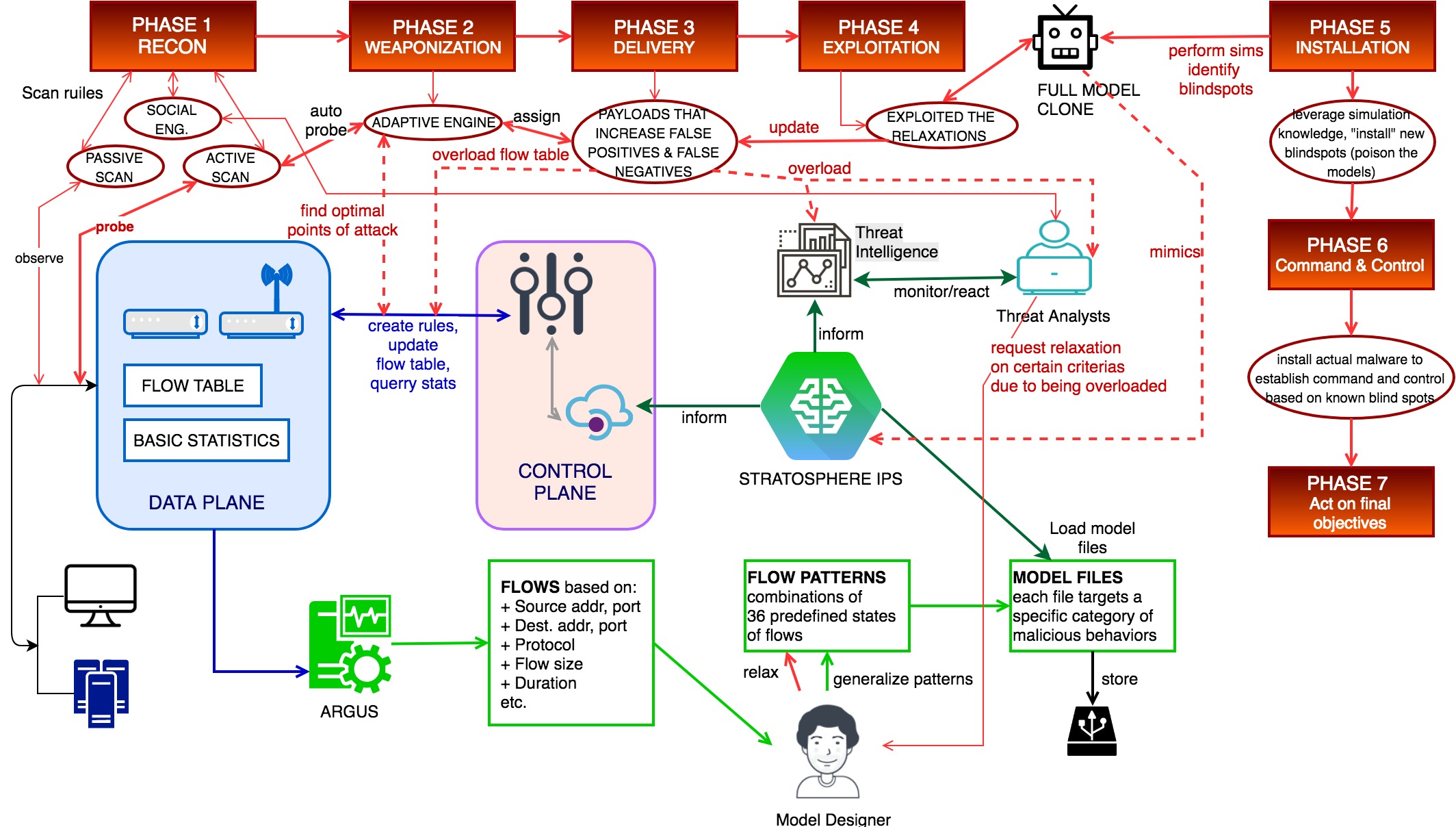}
  \caption{A Sample Kill Chain for Attacking Stratosphere IPS}
  \label{Figure:ThreatModel}
\end{figure*}

\subsection{Equation-solving attack}
This form of attack is fit for logistic regression types such as binary logistic regression (BLR), multi-class logistic regression (MLR), and multi-layer perceptron (MLP). Because the models can be represented as equations with variables, attackers just need to feed the known variable values in, use mathematics to solve the equations and get the rest of the unknown values. For example, with BLR, we have : \newline \quad\quad \(w \in R^{d}, \beta \in R\) with \(f_{i}(x) = \sigma(w \times x + \beta) \) \newline where \newline
\(\sigma(t) = 1/(1+e^{-t})\) \newline
Attacker will feed \(x_{i}\) to the trained model and the model will give \(y_{i} = f(x_{i}) = \sigma(w \times x_{i} + \beta) \). If we have enough \({x_{i},y_{i}}\), we should be able to solve the equations to get \(w\) and \(\beta\). The math will be more complicated when dealing with MLRs and MLPs.With softmax model in MLR for instance, we have:\newline
\(c>2, w \in R^{cd}, \beta \in R^{c}\)\newline
\(f_{i}(x_{i}) = e^{w_{i} \times x + \beta_{i}}/(\sum_{j=0}^{c-1}e^{w_{j} \times x + \beta_{j}})\) \newline
The function can be solved by minimizing its cost/loss functions. The methods proved to be effective in Tramer's experiments \cite{Tramer2016}. With BLR, they were able to achieve \(R_{test}=R_{unif}=0\) with an average probe of 41. The number of probe is much greater with MLR and MLP. For instance,it required them to perform 54,100 queries on average in order to achieve 100\% accuracy on a 20 hidden node MLP. Unlike BLR, it is sometimes very hard to estimate a correct amount of probes needed for MLP and MLR cloning. Especially with MLP, it is hard for attackers to guess how many hidden neuron layers are there, and how many neurons per layer. Attackers will also not be able to tell how many classes an original MLP can identify. However, everything can be different in actual cyber attack scenario. Instead of 100\% accuracy, attackers may only need to clone a model with 90\% accuracy for their purposes, and the amount of probes needed may be significantly lower. Another reason to not aiming for 100\% accuracy is that original ML models get tunned on a fast pace, daily. A 100\% accurate cloned model of today maybe different from the actual model next week.
\subsection{Model inversion attack}
Given feature dimension \(d\) with feature vector \(x_{1},x_{2},...,x_{d}\), some knowledge about some of the features, and access to \(f\) - the model, Fredrikson et al. \cite{Fredrikson2015ModelCountermeasures} proposed that a black box model inversion attack which involves finding an optimal x that maximizes the probability of some known values.\newline
\(x_{opt} = argmax_{x \in X}f_{i}(x)\) \newline
For instance, if an image of Bob was used to train model M to recognize category "man". If that exact image is fed into M, the result will be "man" with 100\% confidence while images do not belong to the training set will never get such absolute score from the model. An attacker can start with one pixel and find the pixel value that gives the maximum score possible of category "man". The process continues to other pixels and the end result is an image very close to Bob's original image in the training set. Tramer et al. \cite{Tramer2016} upgraded this approach by performing inversion attack on a cloned model M' of M. The reported improvement is a 6-hour faster recovering time for 40 faces. This kind of attack opens a theoretical possibility of which attackers can gain some insightful knowledge about a security model's trained data set if they could clone the model with 100\% accuracy and somehow was able to tunnel it out. 
\subsection{Path-finding attack}
Tramer et al. \cite{Tramer2016} also extended prior works on tree attacks and proposed their "path-finding" attack which can be used to map binary trees, multi-nary trees, and regression trees. We have a tree \(T\) with \(v\) nodes and at each node, there is an identifier \(id_{v}\). With \(x \in X \), an oracle query will give \(O(x) = id_{v}\). If \(x \in X_{1} \cup {\perp} \times ... \times X_{d} \cup {\perp}\), \(O(x)\) will return the identifier at the node where T stops. To begin the attack, we pick \(x \in X_{1} \cup {\perp} \times X_{2} \cup {\perp} \times... X_{i1} = [a,b] \times ... \times X_{d} \cup {\perp}\). \(O(x)\) gives \(id_{Lv}\) at the leaves of the tree. We then can separate \([a,b]\) into n sub ranges where n is the number of the corresponding known leaves (at this point). For each \(X_{i1}\) sub range, we repeat the process and find another nodes/leaves. This was referred to as the top-down approach which is of higher performance than the bottom-up approach. Reported performance evaluations of this approach show that in order to achieve 100\% on \(1-R_{test}\) and \(1-R_{unif}\), it will take 29,609 queries to clone a tree with 318 leaves, 8 layers of depth; 1,788 queries to clone a tree with 49 leaves, 11 layers of depth; and 7,390 queries to clone a tree with 155 leaves and 9 layers of depth.
\subsection{Other attacks}
There are several more ways to attack ML models. The Lowd-Meek attack \cite{LowdAdversarialRL} targets linear classifiers that give only class labels as models' outputs. The general idea of this approach is using adaptive queries to throw sample points at the suspected positions of the hyperplane. Another way to attack was described by Bruckner \cite{Bruckner} as a single Stackelberg Prediction Game (SPG). In this game, the Leader (L) is the one with the original ML model M. the Follower (F) is the attacker. F will attack L by generating and feeding model M data that at least will prevent M from learning new knowledge or at most, teach M new faulty knowledge. Theoretically, this can be achieved by providing learning data that maximize the cost function of model M. In real life situations, there are more than one attacker with different attacking goals and L does not know how many F are there and what exactly each F is trying to do. This escalates to the Bayesian Stackelberg Game. Zhou and Kantarcioglu described it as "Nested Stackelberg Game" \cite{Zhou2016ModelingGames} suggesting a solution of using and switching a set of models to confuse the attacker. Kantarcioglu later on also proposed the concept of "planning many steps ahead" in this game. Details of these methods will be further studied and evaluated within a defense-in-depth environment and will be discussed in future works.

\section{Recommendations}

\subsection{Invest time on attack/defense model}
From the beginning, ML scientists should pay attention to the ML cyber kill-chain (the attack model) and at least develop a list of recommendations on safe implementation. Recommendations may include but are not limited to the designer's definition of "attack", the meaning of model's accuracy, the side channels, etc. This is essentially important in the context of open-source. As mentioned before, the definition of "attack" should depend on the model's intended purposes rather than just its accuracy. Some ML based solutions were designed to be multi-purposes. Some solutions were originally designed for a specific purpose but were used for other purposes in real-world implementations. Nonetheless, the designers should clearly communicate the intended purposes of the works they are proposing and ways to protect them. For example, most recent works in ML based security for SDNs have accuracy rates of 98\% (Table \ref{table:MLsolutions}) but the meaning of 0.2\% increase in false negatives may differ greatly from one to another. Based on the attack model, the ML designers may also provide a default protection model, explaining how the structure of their designs fit into the protection model, what may be done to harden their works, what are the security trade-offs to be considered, what are the potential side channels in real world deployments, and so on. While research works are not supposed to be commercial ready, basic recommendations on how to protect and harden a ML based solution by its designers are extremely valuable.

\subsection{Design audit-able model}
ML based solutions should generate meaningful logs or even better, having an interface for the model to be audited automatically. Audits may include information on who made what changes, how much the model has drifted after a period of time, the rates of false positives and false negatives, etc. Ideally, the model itself should be able to give indications on whether or not it is under attack and which stage of the kill chain the attackers are at. ML based solution with good audit capability will also help in case the model needs to be rolled back to its earlier versions.

\subsection{Follow secure development processes}
Because ML based security solutions are softwares, the designers should at least follow a secure software development lifecycle \cite{MicrosoftLifecycle}. It involves secure coding practices, static analysis, test cases, attack surface reviews, and so on. Formal verification is absolutely necessary and should be done to the largest extend possible considering there are huge challenges in performing formal verifications on systems like the artificial neural network. Side channels should be limited and there are mechanisms to protect the privacy of the model and its data. Finally, datasets used for training should be as organic as possible.

\subsection{Design an operational cost model}
Cost is another factor as important as accuracy. For the same purpose, a leaner ML algorithm will usually cost less than a complicated one but there may be cases where it is justifiable to have a complex ML model or even a group of different ML models working together. The designers should at least provide a cost model to make practical sense out of their design decisions. A well designed cost model will help with evaluating the cost of false negatives - a very important metric in ML based solutions for cyber security. The paper "Machine Learning with Operational Costs" from MIT researchers \cite{Tulabandhula2013MachineCosts} may serve as a good start for further readings into optimizing ML operational costs. 

\section{Conclusion}
There is a gap between academic researches on ML based solutions for SDNs and their operational deployments. While research works do not have to meet the strict requirements for a commercial ready product, it is important that solution designers pay attention and establish some initial foundations for the hardening of their works just in case the works are chosen to be implemented in "the wild". It is important to note that there are issues with evaluating the true performance of ML-based SDN security applications and model's accuracy alone will not be enough. Let's not forget that attackers are also equipped with Machine Learning powers and can build systems to predict the behaviors of the defending models. For those reasons, this paper suggests four specific recommendations: \textit{\\1) Pay attention to threat models while designing ML solutions.\\2) Make the ML model audit-able\\3) Follow a secure development process\\4)Produce an initial operational cost model}\\
It is believed that these recommendations will significantly improve the practical properties of ML based solutions for SDNs. Future works will include an automatic system designed to evaluate the robustness of well-known ML based, open source cyber security products such as Apache Spot \cite{Spot2017ApacheSpot}. Hopefully, it can be developed into a threat model assessment tool which can be used to communicate better evaluation metrics of ML-based SDN security solutions

\bibliographystyle{IEEEtran}
\bibliography{IEEEabrv,Mendeley.bib}

\end{document}